\documentclass[aip,amsmath,amssymb, reprint, floatfix, nolongbibliography]{revtex4-1} 

\usepackage{graphicx}

\usepackage[aboveskip=0pt,belowskip=-5pt,font=footnotesize,labelfont=bf]{caption} 
\usepackage{subcaption}
\usepackage{dcolumn}
\usepackage{bm}
\usepackage[version=4]{mhchem} 
\usepackage{braket}
\usepackage[export]{adjustbox} 
\usepackage[section]{placeins}
\usepackage{multirow} 
\usepackage[utf8]{inputenc} 
\usepackage[T1]{fontenc}
\usepackage{threeparttable} 
\graphicspath{{./ci_ccmc}}
\usepackage{cleveref} 
\usepackage{gensymb}
\usepackage{diagbox} 
\usepackage[dvipsnames]{xcolor}
\usepackage{mathtools}
\usepackage{todonotes}
\usepackage{soul}
\DeclarePairedDelimiter \abs{\lvert}{\rvert}%
\newcommand{\T}{\hat T}
\renewcommand{\H}{\hat H}
\newcommand{\angstrom}{\textup{\AA}} 
\usepackage{comment}
\usepackage{array}
\usepackage{cleveref}
\Crefrangeformat{figure}{Fig.~#3#1#4 --~#5#2#6} 
\Crefrangeformat{table}{Tables~#3#1#4 --~#5#2#6} 
\Crefname{figure}{Fig.}{Fig.}
\creflabelformat{equation}{#2#1#3}

\usepackage{chemformula}
\usepackage{simplewick}

\usepackage{mathptmx}
\usepackage{etoolbox}

\makeatletter
\def\@email#1#2{%
 \endgroup
 \patchcmd{\titleblock@produce}
  {\frontmatter@RRAPformat}
  {\frontmatter@RRAPformat{\produce@RRAP{*#1\href{mailto:#2}{#2}}}\frontmatter@RRAPformat}
  {}{}
}%
\makeatother
\begin{document} 
\title[]{A hybrid stochastic configuration interaction--coupled cluster approach for multireference systems}

\author{Maria-Andreea Filip} 
\email{maf63@cam.ac.uk} 
\affiliation{ Yusuf Hamied Department of Chemistry, University of Cambridge,
Cambridge, UK }
\affiliation{Peterhouse, University of Cambridge, Cambridge, UK}
\author{Alex J W Thom} 
\email{ajwt3@cam.ac.uk}

\affiliation{ Yusuf Hamied Department of Chemistry, University of Cambridge,
Cambridge, UK }%

\begin{abstract} 
The development of multireference {coupled} cluster (MRCC) techniques has remained an open area of study in electronic structure theory for decades due to the inherent complexity of expressing a multi-configurational wavefunction in the fundamentally single-reference coupled cluster framework. The recently developed multireference coupled cluster Monte Carlo (mrCCMC) technique uses the formal simplicity of the Monte Carlo approach to Hilbert space quantum chemistry to avoid some of the complexities of conventional MRCC, but there is room for improvement in terms of accuracy and, particularly, computational cost. In this paper we explore the potential of incorporating ideas from conventional MRCC --- namely the treatment of the strongly correlated space in a configuration interaction formalism --- to the mrCCMC framework, leading to a series of methods with increasing relaxation of the reference space in the presence of external amplitudes. These techniques offer new balances of stability and cost against accuracy, as well as a means to better explore and better understand the structure of solutions to the mrCCMC equations.  
\end{abstract} 

\maketitle

\section{Introduction} 

Generally, \textit{ab initio} wavefunction based electronic structure methods aim to include electron correlation effects unaccounted for in a zeroth order wavefunction such as that obtained from a self-consistent field (SCF) approximation. These effects are commonly split into dynamic and static correlation, with largely distinct physical origins.
While the former can be successfully treated by coupled cluster (CC) theory, in particular the ever-popular CCSD(T)\cite{Raghavachari1989} approach, the latter induces a strongly multi-configurational wavefunction, which cannot be easily described by the exponential CC \textit{Ansatz}.

The most commonly used approaches combine different algorithms to solve the static and dynamic correlation problems independently: a complete active space self-consistent field (CASSCF) calculation aims to capture the static correlation in the system, while a many-body perturbation theory (MBPT) calculation is used to include remaining dynamic correlation. Depending on the Hamiltonian partitioning used, this leads to methods like complete active-space second-order perturbation theory (CASPT2)\cite{andersson_second-order_1990} and $n$-electron valence perturbation theory (NEVPT) \cite{angeli_introduction_2001,angeli_n-electron_2001,angeli_n-electron_2002}.

The treatment of strong static correlation using coupled cluster methods remains an area of active research. Numerous  
multireference coupled cluster (MRCC) methods have been developed to tackle this problem, using both state-universal approaches, which compute multiple electronic states of the system simultaneously, and state-specific ones, which target one state 
at a time (see \onlinecite{Lyakh2011} for a review of such methods). Generally, MRCC techniques have not achieved the same popularity as their single-reference counterpart due to various challenges with size-consistency,\cite{Lyakh2011} intruder states{\cite{Kaldor1988,Paldus1993,Jankowski1994,Paldus1994a} }and the higher system-specific knowledge requirement to use them effectively.
An alternative approach has been to use externally corrected coupled cluster methods,\cite{paldus_externally_2017} which make use of approximations to higher order cluster amplitudes from {other} sources such as adaptive configuration interaction (ACI)\cite{schriber_communication_2016} or full configuration interaction quantum Monte Carlo (FCIQMC) and coupled cluster Monte Carlo (CCMC).\cite{deustua_converging_2017,deustua_communication_2018,deustua_high-level_2021} Finally, the tailored coupled cluster methods use amplitudes from {another} method, such as CASCI\cite{kinoshita_coupled-cluster_2005,morchen_tailored_2020,vitale_fciqmc-tailored_2020} or the density matrix renormalization group (DMRG) method\cite{Veis2016}, to include non-dynamical correlation into a single-reference coupled cluster calculation, during which these {initial} amplitudes are kept fixed. 

FCIQMC\cite{Booth2009} and CCMC\cite{Thom2010} are stochastic alternatives to traditional \textit{ab initio} 
methods which have been 
developed in the last decade to take advantage of the sparsity present in most electronic Hamiltonians 
to significantly decrease calculation memory requirements. The CCMC algorithm also permits easy generalisation to 
large cluster truncation levels,\cite{Neufeld2017} 
which allows high-accuracy calculations for systems with significant 
contributions from highly excited determinants. While simple to implement, 
these methods can still be computationally prohibitive, scaling as $\mathcal{O} (N^{2n+2})$,
where $N$ is a measure of system size and $n$ is the cluster expansion truncation level. 
We recently developed a state-specific multireference CCMC\cite{Filip2019} approach that 
is highly flexible, using an 
arbitrary reference space and potentially distinct truncation levels with 
respect to each reference. These properties can be used to define 
optimised calculations but doing so requires careful advance investigation of 
the system's Hilbert space, which can be time- and resource-consuming for large 
systems. A common approach employed in MRCC methods is to use a complete active 
space (CAS) as a reference space and initialise it with a high accuracy 
method such as CASCI or CASSCF. While this is not necessarily always the 
optimal reference, it requires minimal system-specific knowledge to implement.
 
In this paper, we investigate ways of combining a configuration
interaction quantum Monte Carlo (CIQMC) calculation in a CAS with a 
multireference CCMC calculation to treat the external space. 
We develop schemes allowing increasing relaxation of the CASCIQMC wavefunction in the
presence of the external mr-CCMC contributions and explore their properties using a
minimal basis square H$_4$ model, as well as the 4- and 8-site 2D Hubbard model.
We also use a range of toy Li systems to explore the performance of each
approach as the number of core and virtual orbitals increases.

\section{Quantum Monte Carlo Methods} 

\subsection{Full Configuration Interaction Quantum Monte Carlo} 
FCIQMC\cite{Booth2009} encodes a stochastic solution to the FCI equations.
The FCI wavefunction is expressed as a linear combination of Slater determinants,
generally obtained by considering all possible arrangements of the desired number
of electrons among spin-orbitals obtained from a Hartree--Fock (HF) calculation.
\begin{equation} 
\ket{\Psi_\mathrm{FCI}} = (1 + \hat C)\ket{D_0}
\end{equation}
where $\ket{D_0}$ is the Hartree--Fock reference determinant, $\hat C = \sum_{i,a} c_i^a \hat a_i^a +
\frac{1}{4}\sum_{i,j,a,b}c_{ij}^{ab} \hat a_{ij}^{ab} + ... $ and $\hat a_i^a,
\hat a_{ij}^{ab}$ are excitation operators, exciting electrons from spin-orbitals $i, j,...$ to spin-orbitals $a, b, ...$.  

{FCIQMC belongs to the family of Projector Monte Carlo (PMC) methods, in which the ground state wavefunction is obtained by solving the imaginary time Schr\"odinger equation,

\begin{equation}
    \frac{\partial \ket{\Psi}}{\partial \tau} = - \hat H \ket{\Psi},
    \label{eq:im_sch}
\end{equation}
where $\tau$ is imaginary time and $\hat H$ is the Hamiltonian of the system of interest.
This has a solution of the form
\begin{equation}
    \ket{\Psi_0} = \lim_{\tau \rightarrow \infty} e^{-\tau (\hat H - E_0)}\ket{\Psi(0)},
    \label{eq:lim}
\end{equation}
where $\ket{\Psi(0)}$ is some initial trial wavefunction such that $\braket{\Psi_0|\Psi(0)}\neq0$ and $E_0$ is the lowest eigenvalue of $\hat H$. The imaginary-time interval can be split into a series of small time-steps $\delta \tau$ and \cref{eq:lim} can be rewritten in an iterative form as 
\begin{equation}
    \ket{\Psi(\tau + \delta \tau)} = e^{-\delta\tau (\hat H - E_0)}\ket{\Psi(\tau)} \approx (1-\delta\tau(\hat H - E_0))\ket{\Psi(\tau)}.
\end{equation}
If $\ket{\Psi(\tau)}$ is an FCI wavefunction, projection of the equation above onto all the determinants in the Hilbert space, $\bra{D_\mu} = \bra{D_0}(\hat a_{ij...}^{ab...})^\dagger$, gives an iterative update equation for the CI coefficients,
\begin{widetext}
\begin{equation} 
c_\mu(\tau + \delta \tau) = c_\mu(\tau)
\underbrace{- \vphantom{\sum_\nu}\delta\tau \braket{D_\mu|\hat H -E_0|D_\mu}c_\mu(\tau)}_{\mathrm{Death}} \underbrace{-
\sum_{\nu}\delta\tau\braket{D_\mu|\hat H|D_\nu}c_\nu(\tau)}_{\mathrm{Spawning}}.
\label{eq:iter}
\end{equation} 
\end{widetext}}
The CI coefficients on each determinant can be viewed as populations of psips (`psi particles' or `walkers')
residing in the Hilbert space of the system. \Cref{eq:iter} can be solved
stochastically by sampling the population dynamics of the psips as they undergo the following processes:
{ \begin{itemize} 
\item spawning from $\ket{D_\nu}$ to $\ket{D_\mu}$
creates a new particle with probability 
\begin{equation}
p_\mathrm{spawn}(\nu|\mu) \propto \delta \tau|H_\mathbf{\mu \nu}|;
\end{equation} 
\item death or cloning of particles on $\ket{D_\mu}$ occurs with probability 
\begin{equation}
p_\mathrm{death}(\mu) \propto \delta \tau |H_ \mathbf{\mu\mu} - S|,
\end{equation} 
where the parameter $S$, known as the shift, has been introduced in place of the unknown ground state energy $E_0$.
\item annihilation of particles of opposite sign on the same
determinant. 
\end{itemize} 
The shift in the death step is allowed to vary, acting as a population control parameter and, once
the system has reached a steady state, converges to the true energy $E_0$. In a 
standard FCIQMC calculation, two independent estimators can therefore be used for the
energy -- the shift described above and the projected energy
\begin{equation}
E_\mathrm{proj} = \frac{\braket{D_0|\hat H|\Psi}}{\braket{D_0|\Psi}} =
\sum_{\mu \neq 0} \frac{N_\mu H_{\mu 0}}{N_0},
\end{equation} 
where $N_\mu$ is the FCIQMC population on determinant $\mu$, which should on average be proportional to $c_\mu$. Effective techniques to obtain statistical estimates of these quantities from the QMC imaginary-time series have been devised\cite{Ichibha2022} and for large enough particle populations they are unbiased estimators of the true energy. For low populations, a bias is observed but can be minimised by using appropriate calculation parameters.\cite{Vigor2015}}
\subsection{Coupled Cluster Monte Carlo} 
\label{sec:ccmc} 

{ Rather than a CI expansion, one can use a CC exponential \textit{Ansatz} for the QMC trial wavefunction.}
\begin{equation} 
\ket{\Psi_\mathrm{CC}} = e^{\hat T} \ket{D_0} 
\end{equation}
where $\hat T = \sum_{i,a} t_i^a \hat a_i^a +
\frac{1}{4}\sum_{i,j,a,b}t_{ij}^{ab} \hat a_{ij}^{ab} + ... $.  

When the system is dominated by dynamic correlation, the HF reference is a good
first-order approximation to the true wavefunction, so we expect the cluster amplitudes
$\{\mathbf{t}\}$ to be small. {Hence
$\braket{D_\mu|\Psi_\mathrm{CC}} = t_\mu + \mathcal{O}(t^2)$ and we can write
\begin{equation} 
t_\mu(\tau) - \delta\tau \braket{D_\mu|\hat H -E_0|\Psi_\mathrm{CC}} \approx t_\mu(\tau + \delta
\tau).  
\label{eq:pop_dyn} 
\end{equation} }
This equation can be described by the same population dynamics considered for FCIQMC, { using once again a variable shift to estimate the unknown ground state energy $E_0$,
but one must take into consideration
contributions from ``composite clusters", where we define a cluster to be any combination of excitation operators that appears in the expansion of the coupled cluster exponential Ansatz.  For example, }
\begin{equation}
\braket{D_{ij}^{ab}|\Psi_\mathrm{CC}} = \underbrace{t_{ij}^{ab}}_\mathrm{non-composite\ cluster} + \underbrace{t_i^at_j^b - t_i^b t_j^a}_\mathrm{composite\ clusters},
\end{equation} 
and therefore any of these three terms may contribute to death on
$t_{ij}^{ab}$ or to spawning onto some excitor coupled to it by the Hamiltonian.
{The selection process for terms in \cref{eq:pop_dyn} is therefore somewhat more complicated than for FCIQMC. }
The original selection algorithm begins by selecting a cluster size $s$ with 
probability 
\begin{equation} 
p(s) = 
\begin{cases}
\frac{1}{2^{s+1}}, & s < s_\mathrm{max} \\
\frac{1}{2^{s}}, & s = s_\mathrm{max}.
\end{cases}
\end{equation}

A particular cluster of $s$ distinct {excitation operators or excitors} is then selected with probability
\begin{equation}
p(e|s) = s! \prod_{i=1}^s \frac{|N_i|}{|N_\mathrm{ex}|}
\label{eq:psel} 
\end{equation} 
where $N_\mathrm{ex}$ is the total population on
excitors. The total selection probability is therefore 
\begin{equation}
p_\mathrm{sel}(e) = p(e|s)p(s) = 
\frac{s! \prod_{i=1}^s \frac{|N_i|}{|N_\mathrm{ex}|} }{2^{s+1}}
\end{equation}

{ Once this cluster is selected, the determinant obtained by applying it to the reference determinant must be computed, so that it may be used in the stochastic processes described previously. However, care must be taken to account for potential sign differences between the cluster representation of a determinant, $\prod_e \hat a_e \ket{D_0}$ and the determinant itself $\ket{D_{ij...}^{ab...}}$. due to the commutation properties of the excitation operators.}

Various improvements have been developed for the CCMC algorithm,\cite{Franklin2016, Spencer2016,Neufeld2018, Spencer2018} including an
importance-sampling-based selection method, which avoids the disproportionately
high likelihood of selection of unimportant large clusters in the original method.\cite{Scott2017}

\subsection{Multireference Coupled Cluster Monte Carlo} 

For a multireference coupled cluster approach, the Hilbert space can 
be partitioned into a reference (model) space and an external
space. One approach, first suggested in the work or Piecuch, Oliphant and Adamowicz,\cite{Oliphant1991,Oliphant1992,Piecuch1993, Piecuch1994} uses a formally single-reference
\textit{Ansatz}, 
\begin{equation} \ket{\Psi_\mathrm{CC}} = e^{\hat T_\mathrm{int} + 
\hat T_\mathrm{ext}}\ket{D_0}, 
\label{eq:Ansatz}
\end{equation} 
where $\hat T_\mathrm{int}$ excite electrons within the reference space while 
$\hat T_\mathrm{ext}$ includes at least one external excitation.  Given a principal reference
$\ket{D_0}$ each other 
determinant in the reference space (referred to as `secondary references' in the following) can be expressed as 
{$\ket{D_n} = \pm \hat a_n \ket{D_0}$}, where $\hat a_n$ is some excitation operator, 
we can also express a multireference coupled cluster wavefunction as 
\begin{equation} 
\ket{\Psi_{\mathrm {CC}}}=
\mathrm{exp}[\sum_{i=1}^{m_0} \hat T_{i} + 
\sum_{n=1}^{N}\sum_{j=0}^{m_n}\hat T^{(n)}_{j}t_n\hat a_n]\ket{D_{0}}, 
\label{eq:mrcc} 
\end{equation} 
where $\hat T_i$ are {external} $i$-th order excitors of the first reference, $\hat T^{(n)}_j$ are {external} $j$-th
order excitors of the $n$-th secondary reference, $m_n$ is the truncation
level associated with each reference { and $t_m$ are the coefficients of the internal excitation operators}. This form allows for a completely general
reference space and set of excitation levels to be used. However, some external
determinants may be within $m_n$ excitations of multiple references, so care
must be taken to only include the relevant excitors in one $\hat T^{(n)}$. Writing
all excitations in terms of a single reference circumvents this problem, but the
form in \Cref{eq:mrcc} is helpful in understanding the corresponding
stochastic algorithm.

In the $m$-reference CCMC ($m$r-CCMC) aproach,\cite{Filip2019} all but the main reference 
$\ket{D_0}$ are used to define an appropriate selection space. In particular, 
one must compute $n_\mathrm{max} =\max\limits_{n}\{e_n + m_n\}$, 
where $e_{n}$ is the excitation level of reference $n$ with respect to 
$\ket{D_0}$. Then, clusters of size up to $n_\mathrm{max} + 2$ are selected,
but only those within $m_n +2$ of at least one
reference are accepted. Spawning, death and annihilation processes are
unchanged, as is the computation of the shift and projected energy. 

\section{Multireference CCMC based on a CASCIQMC wavefunction}
\label{sec:embedded} 
\subsection{General considerations}
\label{sec:gen}
In developing an mr-CCMC method based on a {CASCIQMC} reference wavefunction, 
we consider the {Jeziorski-Monkhorst} \textit{Ansatz},\cite{Jeziorski1981} given by
\begin{equation} 
 \ket{\Psi_\nu} = \sum_\mu c_{\nu\mu} e^{\hat T^\mu},
\ket{D_\mu} 
\end{equation}
{where $\ket{\Psi_\nu}$ correspond to different approximate eigenstates of the Hamiltonian, $c_{\nu\mu}$ are the CI coefficients and $e^{\hat T^\mu}$ the external cluster operators corresponding to reference wavefunction $\ket{D_\mu}$. While this \textit{Ansatz} was originally used in state-universal methods,\cite{Paldus1993,Paldus1994a} a state-specific approach known as Mk-MRCC was developed by Mahapatra \textit{et al.}\cite{Mahapatra1998,Mahapatra1999}} In this method, obtaining $c_\mu$ and $\hat T^\mu$ requires the
simultaneous optimisation of a CI wavefunction within the CAS reference space

{ As originally noted in \onlinecite{Piecuch1993, Piecuch1994}, the zeroth order reference-space part of the Jeziorski-Monkhorst Ansatz is equivalent to the corresponding part of the mr-CCMC Ansatz, if one re-expresses the
reference space wavefunction in terms of  a CI expansion rather than an internal cluster operator,}
\begin{equation}
    e^{\hat T_\mathrm{int}}\ket{D_0} \approx \sum_\mu c_\mu \ket{D_\mu},
\end{equation}
{ This approach is also taken in the semi-linearised CASCC method,\cite{Adamowicz1998, Ivanov2000, Ivanov2009} which employs this transformation on the operator in \cref{eq:Ansatz} and a CAS reference space to successfully treat multiconfigurational problems.}

\begin{figure}[ht] 
\centering
\includegraphics[width =0.5\textwidth, trim = 0 0 0 1.5cm, clip]{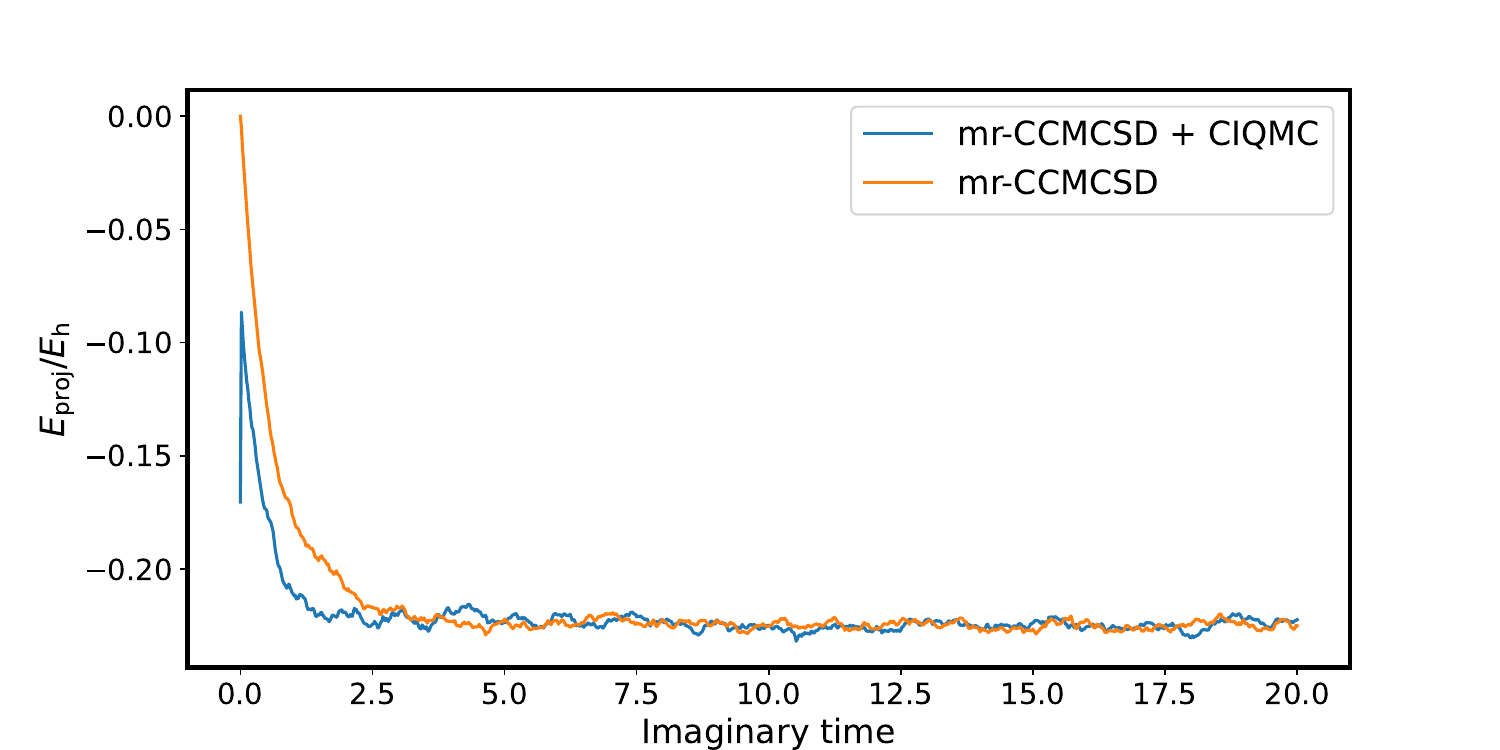} 
\caption[Convergence of mr-CCMC $E_\mathrm{proj}$ initialised from CASCIQMC.] {\protect \footnotesize \raggedright Convergence of $E_\mathrm{proj}$ in mr-CCMC calculations initialised from CASCIQMC or 
from the HF wavefunction for N$_2$ at $r_\mathrm{NN} = 1.3$ \AA in the STO-3G basis, { using the standard mr-CCMC algorithm.} }
\label{fig:N2-cicc} 
\end{figure}

It is therefore worth considering whether this approach to solving for the reference-space wavefunction may bring any benefits over the cluster expansion approach { in the stochastic paradigm}. While the two representations
are largely interchangeable, sampling of a CI wavefunction is more straightforward than that of the corresponding CC wavefunction, making the former a potentially
faster route to generating a CAS wavefunction, which can be then used as a starting point for a full-space mr-CCMC
calculation. This initialisation alone leads to faster convergence of the mr-CCMC calculation, as can be seen in \Cref{fig:N2-cicc}.

We describe a pair of approximations to the full mr-CCMC method which use the CASCIQMC wavefunction as a starting point.
In the first frozen-reference mr-CCMC (fr-mr-CCMC) approach, no relaxation of the reference wavefunction from its CASCIQMC value is allowed. { This is the same concept employed in TCC,\cite{kinoshita_coupled-cluster_2005} although the full cluster expansion of the CAS wavefunction is preserved, rather than maintaining only the $\hat T_1$ and $\hat T_2$ terms.} This can be 
achieved by only permitting death and spawning into the external space. A second approach, referred to as relaxed-reference (rr-mr-CCMC) allows relaxation of 
the reference space in response to the external clusters, but not to internal changes. This corresponds to allowing spawning into the reference space from the external space, but no internal-internal spawning.

\subsection{Combining CIQMC and CCMC populations}

Implementing any of the approaches discussed above involves a mixture of CIQMC and CCMC amplitudes, { which will have different numerical values for an arbitrary determinant, when used to describe the same wavefunction.}
The simplest approach is to translate the CASCIQMC wavefunction into a CCMC form, by a process known as cluster analysis. Alternatively, the mixed CI-CC representation can be maintained and the selection algorithm 
altered to account for this. 

\subsubsection{Cluster analysis}

Consider expressing a wavefunction in two equivalent ways:
\begin{equation}
\ket{\Psi} = (1 + \hat C)\ket{D_0} = e^{\T} \ket{D_0}.
\end{equation}
Expanding each of these expressions gives
\begin{equation}
\ket{\Psi} = (1 + \sum_{ia} c_i^a \hat a^\dagger \hat i + \sum_{ijab} c_{ij}^{ab} \hat a^\dagger \hat b^\dagger \hat j \hat i + ...) \ket{D_0}
\end{equation}
and
\begin{equation}
\ket{\Psi} = (1 + \sum_{ia} t_i^a \hat a^\dagger \hat i + \sum_{ijab} (t_{ij}^{ab} - \frac{t_i^at_j^b}{2!} + \frac{t_i^b t_j^a}{2!}) \hat a^\dagger \hat b^\dagger \hat j \hat i + ...) \ket{D_0}
\end{equation}
respectively.
Equating each term from the sums gives equations of the form
\begin{equation}
\begin{split}
&c_i^a = t_i^a\\
&c_{ij}^{ab} = t_{ij}^{ab} - \frac{t_i^at_j^b}{2} + \frac{t_i^b t_j^a}{2}\\
&\mathrm{etc.}
\end{split}
\end{equation}
Given an intermediately normalised CI wavefunction, these equations can be used to construct the corresponding CC expansion. 
In the context of QMC methods, the normalisation of the wavefunction must be accounted for. The easiest way to maintain this consistently
is to divide the obtained CIQMC populations by $N_0$, perform the cluster analysis and re-multiply the resulting CC amplitudes by the same $N_0$. This 
also provides an easy means to modify the normalisation between the CIQMC and mr-CCMC calculations if desirable.
\subsubsection{Mixed CI-CC sampling}
If a CI wavefunction is preserved in the reference space, the \textit{Ansatz} has the form
\begin{equation}
\ket{\Psi} = e^{\T_\mathrm{ext}}(1 + \hat C_\mathrm{int})\ket{D_0}.
\end{equation}
Given the form of this expansion,  each { cluster of excitors selected in CCMC as described in \cref{sec:ccmc} must contain a contribution
from up to one internal excitation operator, as $\hat C_\mathrm{int}$ is a sum of these. Additionally, as the excitation operators commute and the internal and external operator pools are distinct, one can choose to always apply the internal excitation first when working out the determinant corresponding to a particular CCMC cluster selection.} It is therefore possible to sample this expansion in the following way:
\begin{enumerate}
\item Select a cluster size $s$ with some probability $p_s$. This can be done with any of the available selection schemes.
\item Select a reference with probability
\begin{equation}
p(r) = \frac{\abs{N_r}}{\sum_{i \in \mathrm{refs}}\abs{N_i}}.
\end{equation}
\item If the selected reference is $\ket{D_0}$, then select $s$ excitors with
\begin{equation}
p(e|s) = s!\prod_{i=1}^s \frac{|N_i|}{N_\mathrm{ex}},
\end{equation}
where $N_\mathrm{ex}$ is now the total population on non-reference excitors. If the selected reference is not $\ket{D_0}$, then $s-1$ excitors are selected with 
\begin{equation}
p(e|s) = (s-1)!\prod_{i=1}^{s-1} \frac{|N_i|}{N_\mathrm{ex}}.
\end{equation}
\end{enumerate}
The rest of the algorithm is largely left unchanged, although care must be taken to account for potential sign differences between { $\ket{D_{ij...}^{ab...}}$ and $\hat a_{ij...}^{ab...}\ket{D_0}$}.

\section{Approximate mr-CCMC}
\subsection{ Perturbation theory treatment}
\label{sec:math}
We characterise the proposed approximate mr-CCMC approaches mathematically using small
example systems and the framework of perturbation theory (PT). This allows us to
predict some of the properties and limitations of these methods.
\newline\newline
Consider first a system with two reference determinants, $\ket{D_0}$ and
$\ket{D_1}$, and one external determinant, $\ket{D_2}$. The CIQMC equations in
the reference space are given by 
\begin{align} 
\begin{split} 
\braket{D_0| \hat H - S_\mathrm{ref}|\Psi} = 0\\ 
\braket{D_1|\hat H - S_\mathrm{ref}|\Psi} = 0, 
\end{split} 
\end{align}
{ where we have substituted the shift $S_\mathrm{ref}$ for the unknown ground state energy of the reference space. These expand to}
\begin{align} 
\begin{split} (H_{00} - S_\mathrm{ref})c_0 + H_{01}c_1 = 0 \\
(H_{11} - S_\mathrm{ref})c_1 + H_{10}c_0 = 0. 
\end{split} 
\end{align} 
Setting $H_{00} = 0$ and $c_0 = 1$ gives 
\begin{equation} S_\mathrm{ref} = 
\frac{H_{11} \pm \sqrt{H_{11}^2 +4H_{01}^2}}{2}\ \ \mathrm{and} \ \ 
c_1 = \frac{S_\mathrm{ref}}{H_{01}}. 
\end{equation} 
Outside the reference space, the cluster amplitude equation is 
\begin{equation} 
(H_{22} - S)t_2 + H_{20}c_0 + H_{21}c_1 = 0, 
\end{equation} 
{ with the shift $S$ now replaces the overall ground state energy, which should in general be different from $S_\mathrm{ref}$ as it takes into account external wavefunction contributions. This gives}
\begin{equation} t_2 = \frac{H_{20} +H_{21}c_1}{S-H_{22}}. 
\label{eq:t2} 
\end{equation} 
We observe that, for fixed
$c_0$ and $c_1$ and any value of $S$ there is a value of $t_2$ that obeys
\cref{eq:t2}. This is equivalent to each value of $S$ generating a different
wavefunction. Therefore, when the reference space wavefunction is frozen, $S$ can no longer be freely used as a population
control parameter and, instead, a physically justified value of $S$ is required.
Two obvious options are available: $S = S_\mathrm{ref}$ and $S =
E_\mathrm{proj}$, the use of which allows us to relate the resulting equations
to terms in Rayleigh--Schr\"odinger\cite{Rayleigh1894, Schrodinger1926} (RS) and Brillouin--Wigner\cite{Brillouin1932, Wigner1935} (BW) perturbation theory (PT)
respectively. 

As an example, consider a system with two references and two
external excitors. We can set up a PT problem with $\hat H = \hat H_0 + \hat V$
where the corresponding matrices are 
\begin{equation} 
\mathbf{H}_0 =
\begin{pmatrix} 
0 & H_{01} & 0 & 0\\ 
H_{10} & H_{11} & 0 & 0\\ 
0 & 0 & H_{22} & 0\\ 
0 & 0 & 0 & H_{33} 
\end{pmatrix} 
\end{equation} 
\begin{equation} 
\mathbf{V}
= \begin{pmatrix} 
0 & 0 & H_{02} & H_{03}\\ 
0 & 0 & H_{12} & H_{13}\\ 
H_{20} & H_{21} & 0 & H_{23}\\ 
H_{30} & H_{31} & H_{32} & 0 
\end{pmatrix}, 
\end{equation}
where {$H_{\mu\nu} = \braket{D_\mu| \hat H |D_\nu}$}. Diagonalising the reference Hamiltonian gives a set of zeroth-order wavefunctions,
\begin{align} 
\begin{split} 
&\ket{\phi_0} = c_0 \ket{D_0} + c_1 \ket{D_1} \\ 
&\ket{\phi_1} = c_1 \ket{D_0} - c_0 \ket{D_1} \\ 
&\ket{\phi_2} = \ket{D_2}\\ &\ket{\phi_3} = \ket{D_3}. 
\end{split} 
\end{align} 
At the first order of RSPT, the energy correction 
and the first order wavefunction contributions { to $\ket{\Psi^{(1)}} = \sum_\mu a_\mu^{(1)}\ket{\phi_\mu}$} are
given by: 
\begin{equation} 
E^{(1)} = \braket{\phi_0|\hat V|\phi_0} = 0 
\end{equation} 
\begin{align} 
&a_1^{(1)} = 0, 
\label{eq:PT1a}\\
&a_2^{(1)} = \frac{H_{20}c_{0} + H_{21}c_1}{E_0^{(0)} - H_{22}},
\label{eq:PT1b}\\
&a_3^{(1)} = \frac{H_{30}c_{0} + H_{31}c_1}{E_0^{(0)} - H_{33}}.  
\label{eq:PT1c}
\end{align} 
The second-order wavefunction { contributions to $\ket{\Psi^{(2}} = \sum_\mu a_\mu^{(2)}\ket{\phi_\mu}$ are given by}
\begin{align}  
&a_1^{(2)} = \frac{a_2^{(1)}\widetilde H_{21} +
a_3^{(1)}\widetilde H_{31}}{E_0^{(0)} - E_1^{(0)}},  
\label{eq:PT2a}\\
&a_2^{(2)} = \frac{a_3^{(1)} H_{23}}{E_0^{(0)} - H_{22}},
\label{eq:PT2b}\\
&a_3^{(2)} = \frac{a_2^{(1)}H_{32}}{E_0^{(0)} - H_{33}}, 
\label{eq:PT2c}
\end{align} 
where $\widetilde H_{ij} = \braket{\phi_i|\hat H | \phi_j}$.

Note that the denominator in the equations above is the death term in QMC, if $S =
E_0^{(0)}$. As FCIQMC gives the exact ground state energy in the reference
space, this is equivalent to $S=S_\mathrm{ref}$. If one were to alternatively
work through the above in a Brillouin--Wigner formalism, all appearances
of $E_0^{(0)}$ in the denominator would be replaced by the total energy $E$.
While this is not known \textit{a priori}, QMC provides an estimate of it in
the form of the instantaneous projected energy. Therefore, BWPT is related to
our approach in the same way as described above, if $S = E_\mathrm{proj}$.

{ In this simple PT picture, the two approximations described in \cref{sec:gen} correspond to
 selectively including some of the terms in \Cref{eq:PT1a,eq:PT1b,eq:PT1c,eq:PT2a,eq:PT2b,eq:PT2c}. If the initial CASCI wavefunction in the reference space is frozen one uses all the first-order terms and the purely external second-order terms $a_2^{(2)}$ and $a_2^{(3)}$.
Adding the mixed internal-external second-order $a_2^{(1)}$ term corresponds to also allowing some relaxation of the reference space.}

While these models are very simple, and notably fail to account for any of the added complexities of dealing with cluster amplitudes rather than a linear expansion, they do suggest some considerations when implementing these algorithms. { We can consider a generic form of  \Cref{eq:PT1a,eq:PT1b,eq:PT1c,eq:PT2a,eq:PT2b,eq:PT2c}, in which the denominators depend on an arbitrary parameter $S$ rather than $E_0^{(0)}$. These equations would have poles when $S$ is
very close to one of the diagonal Hamiltonian elements outside the reference
space, $H_{22}$ or $H_{33}$. This is relatively unlikely to be problematic for $S = E \approx E_\mathrm{proj}$, as
the shift tracks the lowest eigenvalue, which should be lower than any diagonal
Hamiltonian elements. There is a more significant risk of interacting with poles
when $S=S_\mathrm{ref}=E_0^{(0)}$,}
as this value will in general be higher than the ground state and may be close
to other Hamiltonian elements. We find that this is indeed a problem in
the strongly correlated regimes of the systems considered here, for which
$S_\mathrm{ref}$ is above the stability threshold for the equations. 

When considering the additional term in {rr-mr-CCMC} (\cref{eq:PT2a}), we observe that unsurprisingly any spawning into
the reference space (given by ${a_2^{(1)}\widetilde H_{21} +
a_3^{(1)}\widetilde H_{31}}$)  must be accompanied by death in this space (given by $E_0^{(0)} - E_1^{(0)}$) to prevent
 uncontrolled population growth. Secondly, we note that all
spawning towards and death in the reference space \textit{should} occur
onto the excited states of the reference Hamiltonian $\ket{\phi_i}$ rather than onto the reference
determinants $\ket{D_i}$. If one is using CASCIQMC
to find the ground state in the reference space, the wavefunctions and energies
of these excited states are unknown. However, enforcing that any
contribution be orthogonal to the CAS ground state is possible and 
sufficient for a viable propagator, as is discussed in \cref{sec:er}.

\subsection{Algorithmic Details}
In all cases, an FCIQMC calculation is run in a CAS. This {active space} is then used as a reference space for an mr-CCMC calculation, with the
population initialised to the FCIQMC values or a cluster expansion thereof. The constraints placed on the mr-CCMC calculation
to obtain approximate solutions are given below.
\subsubsection{Frozen-reference mr-CCMC}

The {fr-mr-CCMC} approach can be simply implemented by rejecting all spawning and death
attempts in the reference space. The shift can then be fixed to a value or set
to track the instantaneous projected energy. The idea of a population threshold
is no longer relevant, so the shift should start varying or be set to its
final value at the beginning of the mr-CCMC calculation. For simplicity, the CC form of the reference wavefunction
is used in this case.

\subsubsection{Relaxed-reference mr-CCMC}
\label{sec:er}
The {rr-mr-CCMC} method could be implemented na\"ively by allowing death within the
reference space and all spawning attempts from determinants in the external space.
As is shown later in \cref{sec:na\"ive_erds}, this implementation would not
generate a correct propagator. Ideally, we would {express the CAS Hilbert space in the
eigenvector basis of the CAS Hamiltonian} and allow each eigenvector to undergo death and
spawning {independently}. This is not possible when starting from a CIQMC calculation, where
only the lowest eigenvector is known. However, we can enforce that excitations
and death within the reference space occur within the subspace orthogonal to the
CIQMC ground state. This is more easily done by preserving the CI representation of the CAS wavefunction. 

Consider a set of vectors $\{\ket{\Phi_k}\}$ which span the reference space, where $\ket{\Phi_0}$ is the CASCIQMC
wavefunction and the others are an arbitrary set of orthogonal vectors spanning the rest of the space. {Starting from \cref{eq:iter}, for reference determinant $\ket{D_\mu}$,  with the shift substituted for the energy and combining the spawning and death terms into one,
\begin{equation}
c_\mu(\tau + \Delta \tau) = c_\mu(\tau) - \Delta \tau \braket{D_\mu|\H - S|\Psi}
\label{eq:erds}
\end{equation}}
The identity operator can be resolved as
\begin{equation}
\hat I = \sum_{k = 0}^{n_\mathrm{ref} - 1} \ket{\Phi_k}\bra{\Phi_k} + \sum_{\nu = n_\mathrm{ref}}^{N} \ket{D_\nu}\bra{D_\nu}.
\end{equation}
Introducing this on either side of the Hamiltonian in {\cref{eq:erds}} and noting that the reference determinant $\ket{D_\mu}$ is orthogonal to all non-reference determinants,
\begin{equation}
\begin{split}
c_\mu (\tau + \Delta \tau) &= c_\mu(\tau) \\
&- \Delta \tau\big(\sum_{k,l} \braket{D_\mu|\Phi_k}\braket{\Phi_k|\H - S|\Phi_l}\braket{\Phi_l|\Psi} \\
&+ \sum_{k,\nu}\braket{D_\mu|\Phi_k}\braket{\Phi_k|\H|D_\nu}\braket{D_\nu|\Psi}\big).
\end{split}
\end{equation}
If internal spawning and death on the ground reference state is explicitly forbidden, this reduces to
\begin{equation}
\begin{split}
c_\mu (\tau + \Delta \tau) &= c_\mu(\tau) \\
&- \Delta \tau\big(\sum_{k} \braket{D_\mu|\Phi_k}\braket{\Phi_k|\H - S|\Phi_k}\braket{\Phi_k|\Psi} \\
&+ \sum_{k,\nu}\braket{D_\mu|\Phi_k}\braket{\Phi_k|\H|D_\mathbf{j}}\braket{D_\nu|\Psi}\big).
\end{split}
\label{eq:erds1}
\end{equation}
Monte Carlo Sampling this equation can be done in the following way:
\begin{enumerate} 
\item For each Monte Carlo cycle, generate
a random state within the reference space $\ket{\Phi_r} =
\sum_{i\in\mathrm{ref}} c_{r\mu}\ket{D_\mu}$ orthogonal to $\ket{\Phi_0}$. Then compute $E_r
= \braket{\Phi_r|\hat H|\Phi_r}$.
\item For each death attempt in the reference space, death probability on each reference
determinant $D_\mu$ is proportional to $|E_r - S|N_r c_{r\mu}$, where $N_r =
\braket{\Phi_r|\Psi_\mathrm{CC}}$.  
\item For each spawning attempt in the
reference space, the
spawning probability onto a determinant $D_\mu$ is proportional to
$\braket{\Phi_r|\hat H|D_\nu}c_{r\mu}N_\nu$ where $D_\nu$ is the determinant 
spawning originates from and $N_\nu$ is its population.
\end{enumerate} 
\subsection{Shift in frozen-reference mr-CCMC} 
Two physically justified options for the shift in a frozen-reference mr-CCMC calculation have been identified: $S=S_\mathrm{ref}$ and
$S=E_\mathrm{proj,inst}$, where $E_\mathrm{proj,inst}$ is the intantaneous projected energy. We investigate the effect of these choices on {fr-mr-CCMC} using the H$_4$ molecule in a square geometry and a minimal basis set,\cite{Jankowski1980} {(see supplementary material for parameters)} while varying the
side-length $a$. 

Owing to symmetry, this molecule has two degenerate frontier
molecular orbitals, which lead to degenerate lowest-energy single-determinant
wavefunctions $\ket{1^\alpha 1^\beta 2^\alpha 3^\beta}$ and $\ket{1^\alpha
1^\beta 2^\beta 3^\alpha}$. These can be used as references in 2r-CCMCSD,
giving the energies shown in \Cref{tab:H4-2r}. Solving the CASCIQMC problem in
the reference space and then using the {fr-mr-CCMC} algorithm with
$S=E_\mathrm{proj,inst}$ gives very good agreement with the full results, however the same is not true for $S=E_\mathrm{ref}$
(see \Cref{tab:H4-2r}). 
\begin{table*}
\centering
\begin{tabular}{|c|c|c|c|c|c|} 
\hline
$r_\mathrm{HH}$ & \shortstack{FCI \\$E$} &\shortstack{2r-CCMCSD \\$S$} & \shortstack{2r-CCMCSD\\ $E_\mathrm{proj}$} & \shortstack{fr-2r-CCMCSD $E_\mathrm{proj}$ \\($S=E_\mathrm{proj,inst}$)} &  \shortstack{fr-2r-CCMCSD $E_\mathrm{proj}$ \\($S=S_\mathrm{ref}$)}\\ 
\hline 
2.0 &-0.08833198&-0.08838(4) & -0.08833(3) & -0.08844(2)& -0.0897(3)\\ 
2.5 &-0.11429334& -0.11423(3) & -0.11424(4)& -0.11384(3)& -0.1189(4)\\ 
3.0 &-0.14896567& -0.14891(2)& -0.14892(4) & -0.14883(4)& -0.1696(6)\\\
5.0 &-0.34077014& -0.34070(5) & -0.3408(1) &-0.34079(8)& -\\ 
7.0 &-0.46322289& -0.46343(5) & -0.4637(2)& -0.4636(2)& -\\ 
9.0 &-0.51183106& -0.51185(4) & -0.5117(1)& -0.5113(2)& -\\ 
11.0 &-0.53447267& -0.53449(8) & -0.5345(1) & -0.5349(1)& -\\
13.0 &-0.54864903& -0.54858(4) & -0.5487(1)& -0.5483(1)& -\\ 
\hline 
\end{tabular}
\caption[H$_4$ energies computed using 2r-CCMC and its frozen-reference approximation.]{\protect \footnotesize \raggedright Values of the shift $S$ and the projected energy $E_\mathrm{proj}$ of H$_4$ computed using 2r-CCMC and its frozen-reference approximation.}
\label{tab:H4-2r} 
\end{table*}
When the shift is set to equal $S_\mathrm{ref}$, the calculations only converge
for relatively short bond lengths, beyond which the population starts growing
exponentially.  

\begin{figure} 
\centering
\includegraphics[width = 0.5\textwidth, trim = 0 0 0 1.4cm, clip]{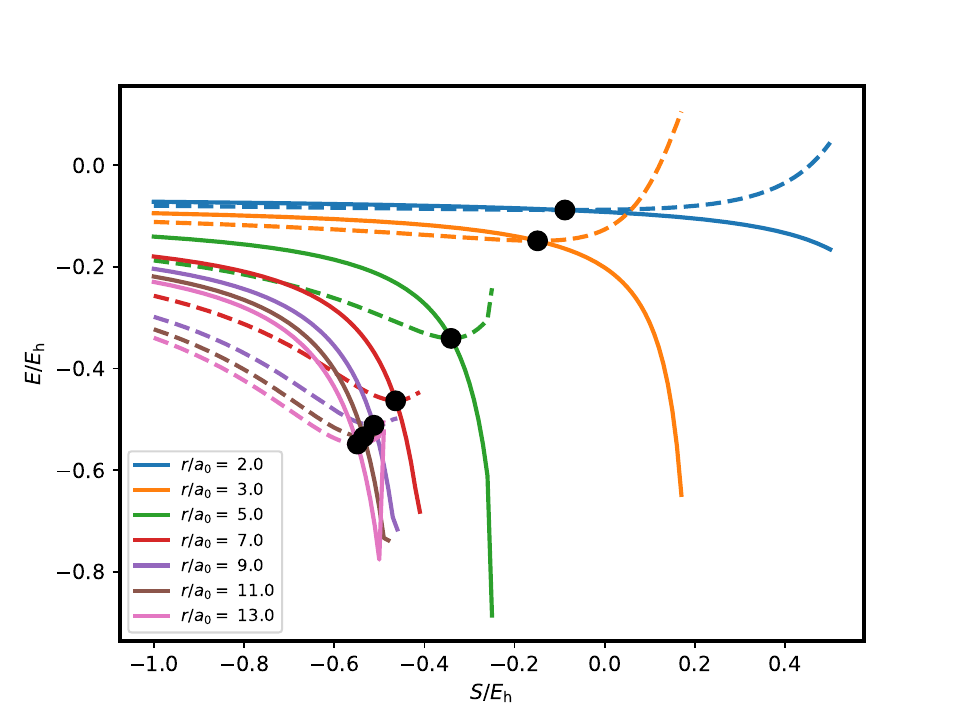}
\caption{\protect \footnotesize \raggedright Converged $E_\mathrm{proj}$ (solid) and variational estimator $\braket{E}$ (dashed) for fr-2r-CCMCSD calculations for
H$_4$ as a function of set shifts, at different values of $r_\mathrm{HH}$. Black circles corresponding to the correct value of $E_\mathrm{proj}$ lie on the $E = S$ line.} 
\label{fig:nrds-evs} 
\end{figure}

{ We prove in \cref{sec:math} that, for a particular model problem, there is 
a solution for any fixed value of the shift. We expect this to be the case in general for this type of ansatz and investigate this numerically here}. A range of such values is considered
in \Cref{fig:nrds-evs}, and the CIQMC and mr-CCMC equations are propagated deterministically
to eliminate any possible uncertainty due to noise. As can be seen, the correct value of the projected
energy can only be achieved in the fixed-shift regime when $S=E_\mathrm{proj}$.
While this point is unremarkable on the projected energy surface, it corresponds
to a minimum in the attained variational energy, as well as a crossing point
between the two surfaces. Therefore, the $S=E_\mathrm{proj, inst}$ approach
corresponds to a variational optimisation of the shift parameter.
\begin{figure*}
\includegraphics[width =\textwidth, trim = 4cm 0 0 1cm, clip]{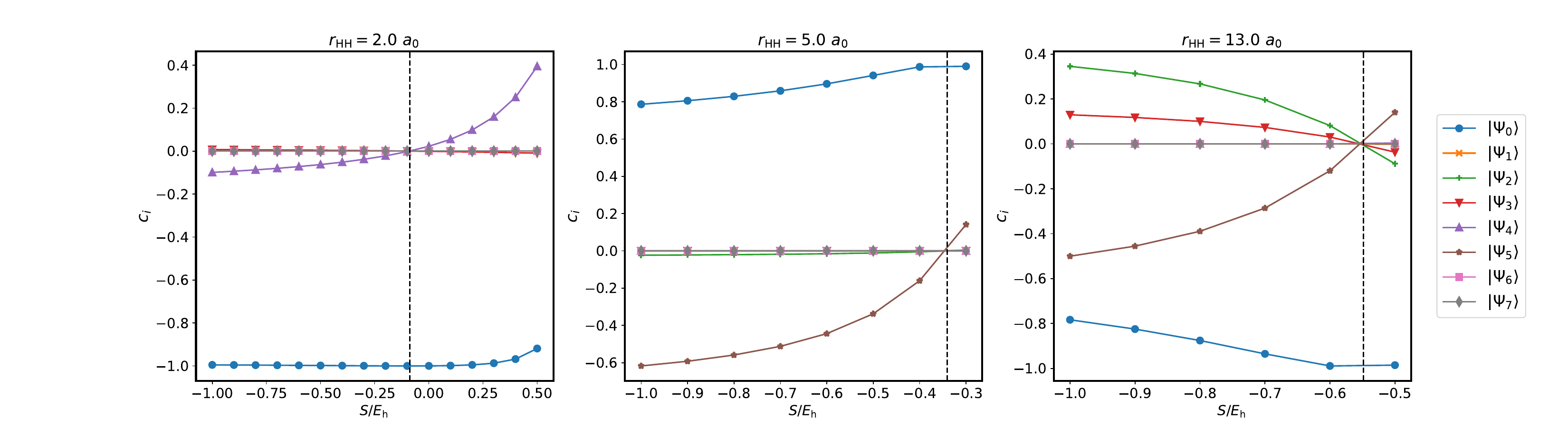}

\includegraphics[width =\textwidth, trim = 4cm 0 0 1cm, clip]{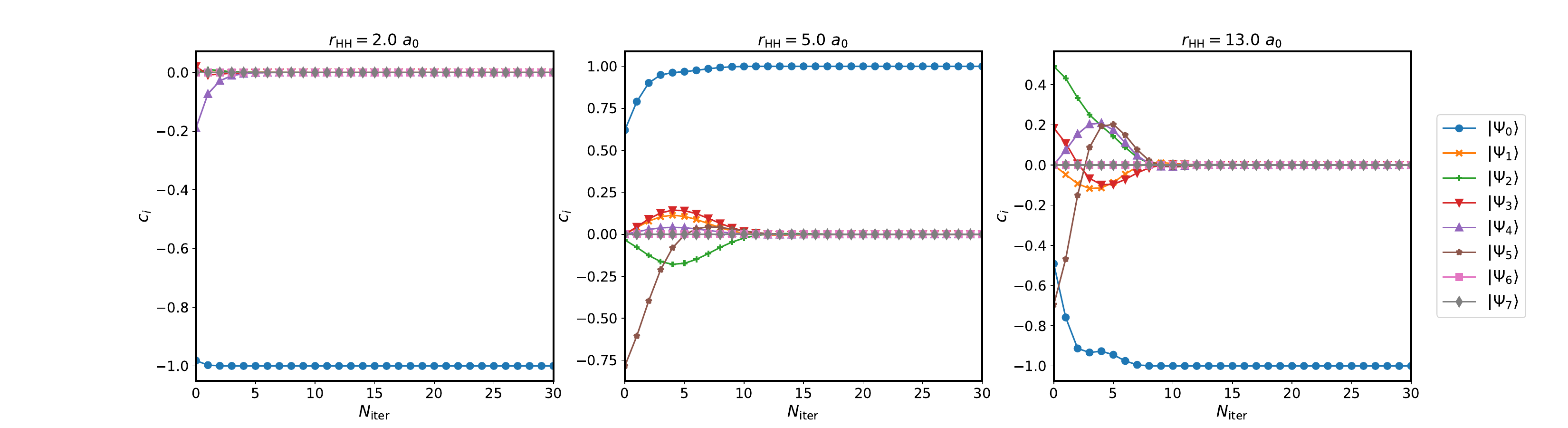}
\caption {\protect \footnotesize \raggedright \textbf{Top:} The projection of the
 final solution of the fr-2r-CCMCSD equations onto the FCI eigenfunctions for various set shifts at H-H distances of
 2.0, 5.0 and 13.0 $a_0$. \textbf{Bottom:} The projection of 
the fr-2r-CCMCSD solution with $S = E_\mathrm{proj,inst}$ onto the FCI eigenfunctions as a function of the number of imaginary 
time steps.}
\label{fig:nrds-projections} 
\end{figure*}  
In this case, the true value of the projected
energy is always within the stability threshold of the calculation, although 
the margin decreases as correlation increases.

We further examine the wavefunctions obtained in the fixed shift regime by
comparing their projections onto the FCI eigenfunctions. In all cases considered in
\Cref{fig:nrds-projections}, the propagation only converges onto a single eigenstate 
at $S=E_\mathrm{proj}$. In comparison, the
$S=E_\mathrm{proj, inst}$ method starts in a mix of states, goes through an
intermediate period where contributions from multiple excited states exist and
quickly settles onto a single eigenstate. Convergence onto a single eigenstate
is possible in this case because frozen-reference 2r-CCMC happens to be equivalent to FCI
for \ce{H4}, but this is not guaranteed in general.

These considerations
suggest the $S=E_\mathrm{proj, inst}$ propagator as the sensible choice for a
fr-mr-CCMC approach, so we will focus on this for further numerical results.
\subsection{The relaxed-reference mr-CCMC propagator}
\label{sec:na\"ive_erds}
\begin{figure*}
\centering
\includegraphics[width =\textwidth, trim = 4cm 0 0 0, clip]{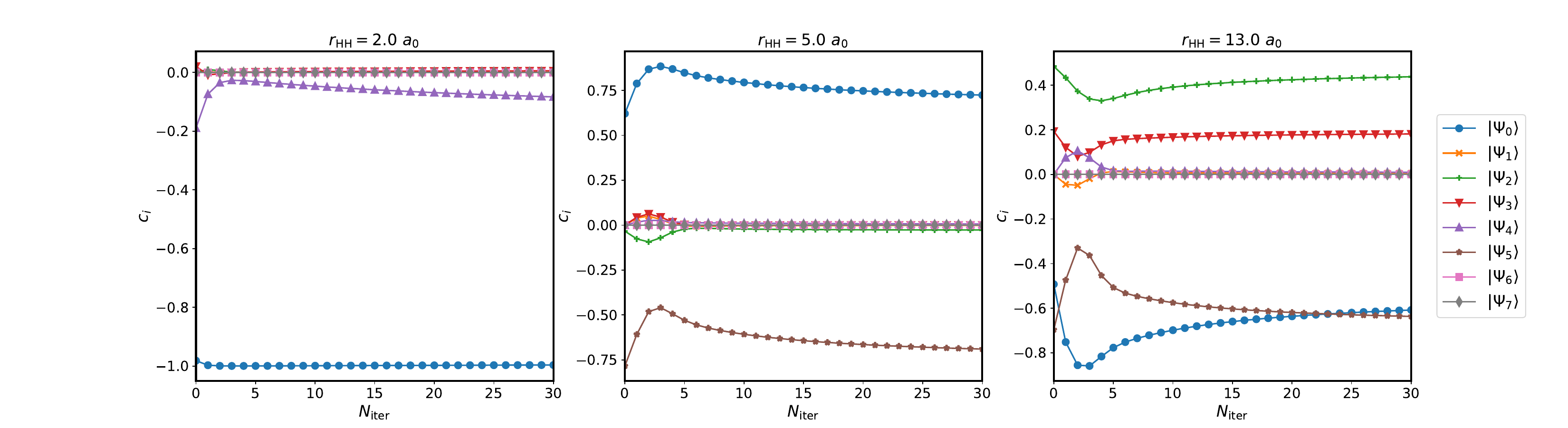}
\caption{ \protect \footnotesize \raggedright Projections onto the FCI eigenfunctions of the na\"ively propagated H$_4$ rr-2r-CCMCSD wavefunction as a 
function of imaginary time at H-H distances of 2.0, 5.0 and 13.0 $a_0$}
\label{fig:erds-projections} 
\end{figure*}

In general, { we expect differences between the CASCI wavefunction and the projection of the true ground state wavefunction onto the CAS. This discrepancy is responsible for systematic errors in TCC,\cite{Faulstich2019} which we expect to appear in fr-mr-CCMC as well. Therefore} we expect methods which allow some relaxation of the
CAS coefficients in the presence of the CCMC wavefunction to give improved
estimates. While not necessary for H$_4$, a correct propagator
which allows external spawning into the reference space should not negatively
impact the quality of the result. As in conventional mr-CCMC, 
calculations set up in this way
have the benefit of a well-defined shift, independent of $E_\mathrm{proj}$,
which can be used for population control. However, the results
of a na\"ive implementation of the rr-mr-CCMC method, which allows independent spawning
and death onto reference determinants, are very poor. The propagator fails to select a single
eigenstate regardless of bond length (see \Cref{fig:erds-projections}). 

\begin{figure*} 
\centering
\includegraphics[width =\textwidth, trim = 4cm 0 0 0, clip]{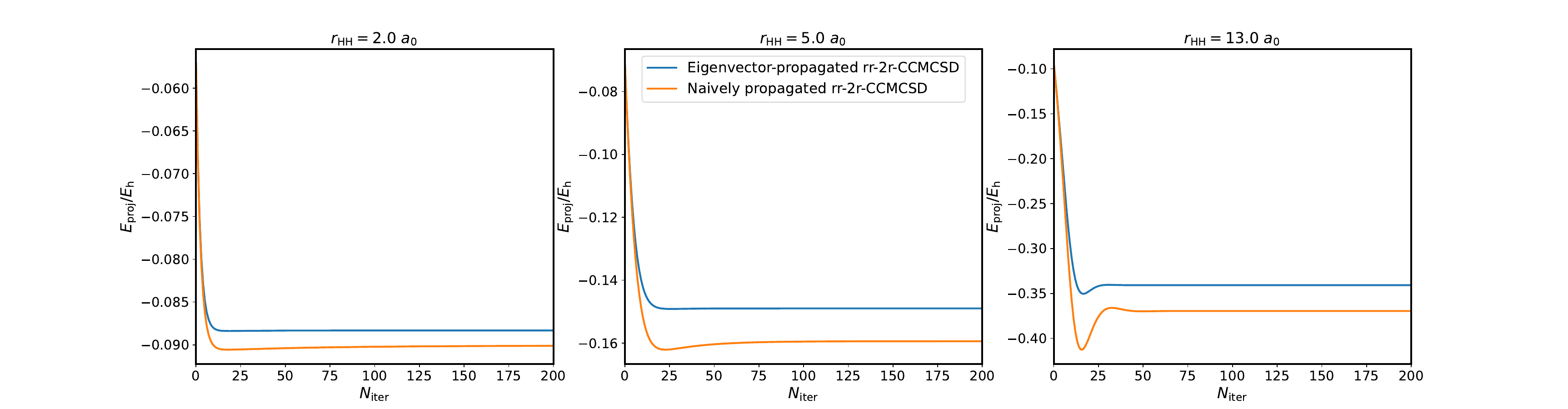} 
\caption{ \protect \footnotesize \raggedright Projected energy of the H$_4$ rr-2r-CCMCSD wavefunction as a function of imaginary 
time at H-H distances of 2.0, 5.0 and 13.0 $a_0$. Naive propagation leads to an overestimate of the correlation
energy.} 
\label{fig:erds-proje} 
\end{figure*}

The failures become increasingly pathological for more correlated geometries. This can be corrected by spawning onto vectors perpendicular to the CASCI wavefunction, as discussed in \cref{sec:er} (see \Cref{fig:erds-proje}). The random orthogonal vector 
propagator is equivalent to the eigenvector-based propagator in this case as there is only one vector
orthogonal to the ground state in the reference space.

In order to meaningfully assess the quality of the random orthogonal vector propagator, we turn to the Hubbard model,\cite{Hubbard1963} used to describe conducting and insulating behaviour in extended lattices. Consider a lattice with sites $r$, each of which has a single spatial orbital centred on it. The Hamiltonian for this system is given by
\begin{equation}
\H = -t \sum_{\braket{r,r'},\sigma} \hat c_{r,\sigma}^\dagger\hat c_{r',\sigma} + U \sum_r \hat n_{r,\uparrow} \hat n_{r, \downarrow}
\end{equation} 
where $\hat c_{r,\sigma}^\dagger$ and $\hat c_{r,\sigma}$ are creation and annihilation operators at site $r$ and $\hat n_{r,\sigma}$ is the corresponding number operator. The pair $\braket{r,r'}$ ranges over adjacent sites.


\begin{figure}
\centering
\includegraphics[width =0.5\textwidth]{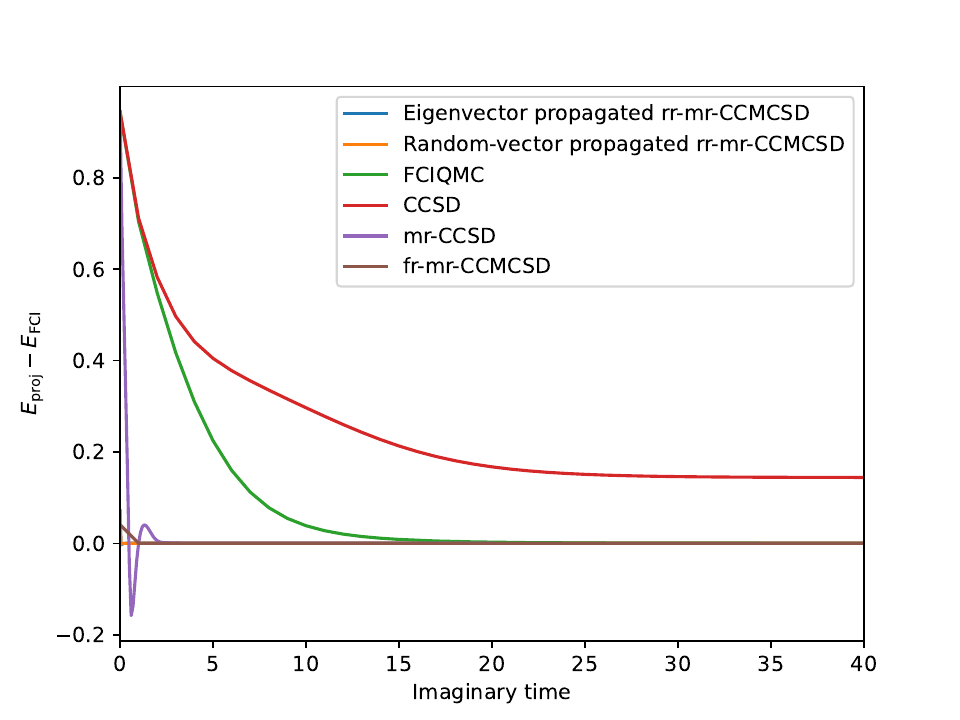} 
\caption[Projected energy as a function
of imaginary time in the 8-site Hubbard model at
$U=t$.]{ \protect \footnotesize \raggedright Projected energy as a function
of imaginary time in the 8-site Hubbard model with  
$U=t$ at $\mathbf{k} = (0, 1)$. The eigenvector propagated rr-mr-CCMCSD line (blue) is obscured by the random-vector propagated
rr-mr-CCMCSD line (orange).} 
\label{fig:h8-2} 
\end{figure}

The hopping integral $t$ and the on-site interaction integral $U$ are free parameters of the model, whose behaviour 
is controlled by the $U/t$ ratio, with higher values corresponding to more strongly correlated systems.
\begin{figure}
    \centering
    \begin{subfigure}{0.22\textwidth}
    \includegraphics[width=\textwidth]{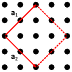}
    \end{subfigure}
    \hfill
    \begin{subfigure}{0.22\textwidth}
    \includegraphics[width=\textwidth]{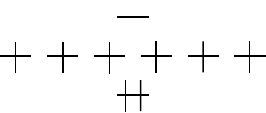}
    \begin{minipage}{.1cm}
            \vfill
            \end{minipage}
    \end{subfigure}
    \caption{Graphic representation of the 8-site Hubbard lattice (left) and the corresponding half-filled MO diagram (right).}
    \label{fig:hub}
\end{figure}
For our investigation we consider the 2D half-filled 8-site Hubbard model described in \Cref{fig:hub}, using momentum space orbitals which represent the 
symmetry conserved HF solutions. Its ground state  falls in the
$\Gamma$-symmetry sector with momentum $\mathbf{k} = (0, 0)$. In this case, the
(6,6)-CASCI ground state corresponds exactly to the CAS contribution to the true
ground state. Moving away from the $\Gamma$ point to the $\mathbf{k} = (0, 1)$ space, two CASCI eigenfunctions
contribute to the true ground state. \Cref{fig:h8-2} gives the energy obtained
by deterministic propagation of FCIQMC as well as single- and multireference CCMC. mr-CCMCSD and rr-mr-CCMCSD with random or true eigenvalue
reference dynamics agree well with each other, while fr-mr-CCMCSD is now slightly above
them in energy --- see \cref{tab:Hubbard}. 
\begin{table}[ht] 
\centering
\begin{tabular}{|c|c|} 
\hline
Method & Projected energy\\ 
\hline 
CCSD  &-0.806154\\ 
48r-CCSD  & -0.949993\\ 
fr-48r-CCSD & -0.949987\\ 
\shortstack{rr-48r-CCSD \\(eigenvector propagation)} & -0.949993 \\ 
\shortstack{rr-48r-CCSD \\(random vector propagation)} & -0.949998 \\ 
FCIQMC & -0.950210 \\ 
\hline 
\end{tabular} 
\caption{\protect \footnotesize \raggedright Converged energies for the 8-site Hubbard model 
with $U$ = $t$ at $\mathbf{k} = (0, 1)$, using different mr-CCMC approaches propagated deterministically.}
\label{tab:Hubbard} 
\end{table} 

\subsection{Stochastic frozen-reference and relaxed-reference mr-CCMC results.}
The previous sections have used deterministic propagations of the mr-CCMC wavefunction and its partially relaxed approximations 
to draw conclusions about these approaches, which we reiterate here. In order to get physically meaningful 
results out of fr-mr-CCMC, the shift must be allowed to exactly follow the projected energy. For rr-mr-CCMC,  
na\"ive propagation is not appropriate, but using random vectors orthogonal to the CASCI ground state as a basis for 
the reference space provides a promising alternative. Making use of these observations, we move to a true stochastic propagation 
of the wavefunction and investigate the quality of these approximations in the presence of random noise for a range of Li-based systems,
which, unlike Hubbard models, have a well-defined core, making it easier to define reference spaces. 

\subsubsection{A highly multiconfigurational species}
\begin{figure} 
\centering
\includegraphics[width = 0.5\textwidth, trim=0.5cm 0 0 1.5cm, clip]{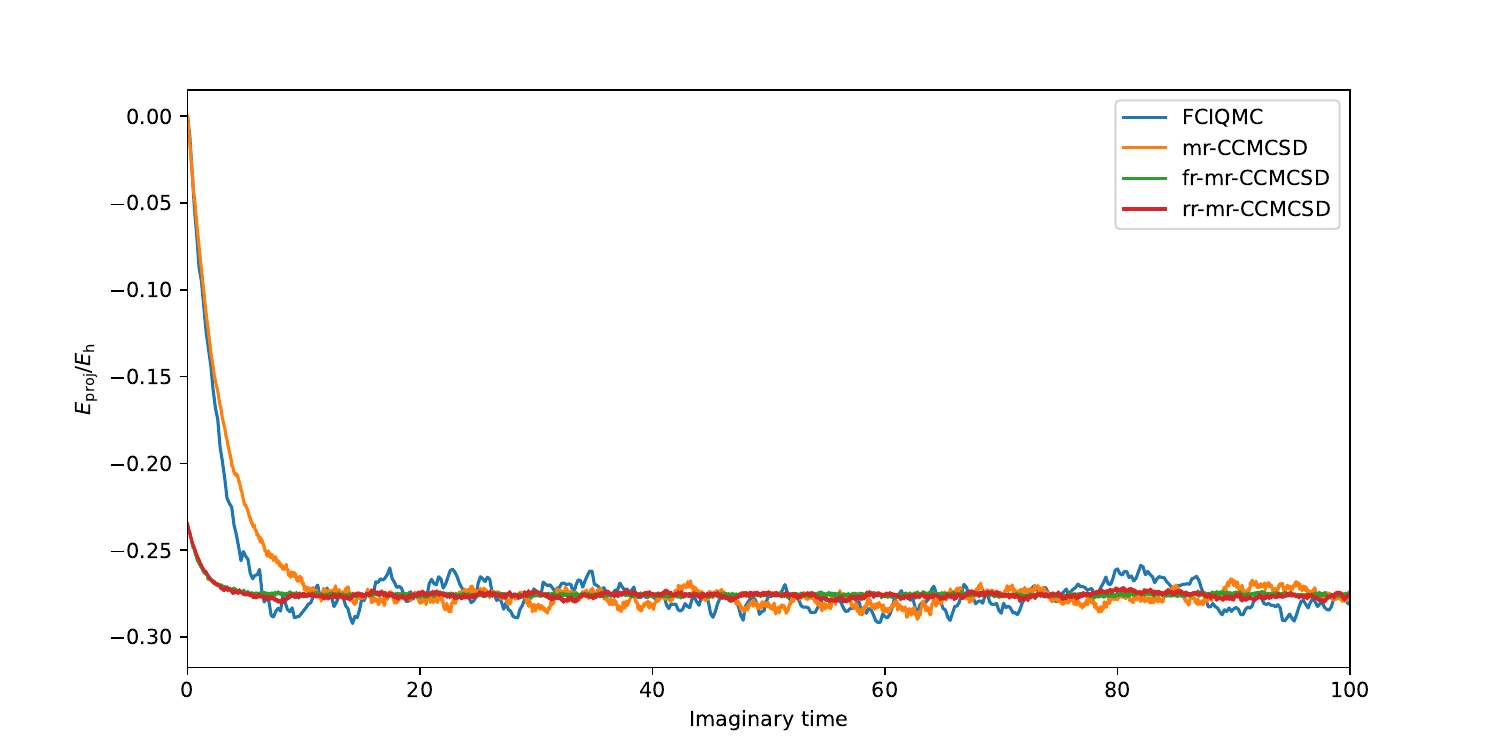}
\caption[Convergence of $E_\mathrm{proj}$ with imaginary time for mr-CCMC in LiH$_3$.]{\protect \footnotesize \raggedright Convergence of $E_\mathrm{proj}$ with imaginary time for FCIQMC, mr-CCMC and the partially relaxed approximations in LiH$_3$. Approximate methods converge faster and lead to lower noise.} 
\label{fig:LiH3} 
\end{figure}

We begin by looking at the \ce{LiH3} molecule in the STO-3G basis. This toy system is designed\footnote{A low-symmetry molecular geometry is used ---
Li at $(0, 0, 0)$ and H atoms at $(4\angstrom, 0, 0)$, $(0, 1\angstrom, 1\angstrom)$ and $(0, -2\angstrom, 1\angstrom)$ --- and orbitals are obtained
from an SCF calculation with an exchange functional that is 10\% Slater--Dirac exchange\cite{Dirac1930} and 90\% Hartree--Fock exchange.} to have many large contributions to the ground state wavefunction, 
dominated by single excitations of the HF wavefunction. The HF wavefunction itself makes a negligible contribution to the ground state, 
making this system particularly challenging for single-reference methods and in this case single-reference CCMCSD fails to converge. 
Using a (4,4)-CAS as a reference space for mr-CCMCSD leads to the results shown in \cref{tab:lih3}.
\begin{table}
\centering
\begin{tabular}{|c|c|} 
\hline
Method & Projected energy\\ 
\hline  
36r-CCMCSD  & -0.2789(6)\\ 
fr-36r-CCMCSD & -0.27566(4)\\ 
rr-36r-CCMCSD  & -0.2760(1) \\ 
FCIQMC & -0.2795(3) \\ 
FCI & -0.279161\\
\hline 
\end{tabular} 
\caption[LiH$_3$ energies from different mr-CCMC approaches.]{\protect \footnotesize \raggedright LiH$_3$ energies from different mr-CCMC approaches, together with FCIQMC and FCI benchmarks.}
\label{tab:lih3} 
\end{table} 
In this case, while full 36r-CCMCSD is within one standard deviation of the FCI value, the approximate methods are not as accurate, although rr-mr-CCMCSD provides an improvement over the frozen-reference approach. However, the approximations are not without benefits. As can be seen in \cref{fig:LiH3}, using either the frozen- or relaxed-reference approach leads to faster convergence and significantly lower noise than the full mr-CCMC or the FCIQMC calculations. 
While the noise is not problematic in this system, CCMC calculations are at times metastable with respect to non-physical solutions with large oscillations likely to push them towards these undesirable alternate solutions. Noise reduction is therefore a valuable property.

\subsubsection{Li structures}
\begin{figure*}
\includegraphics[width =0.45\textwidth]{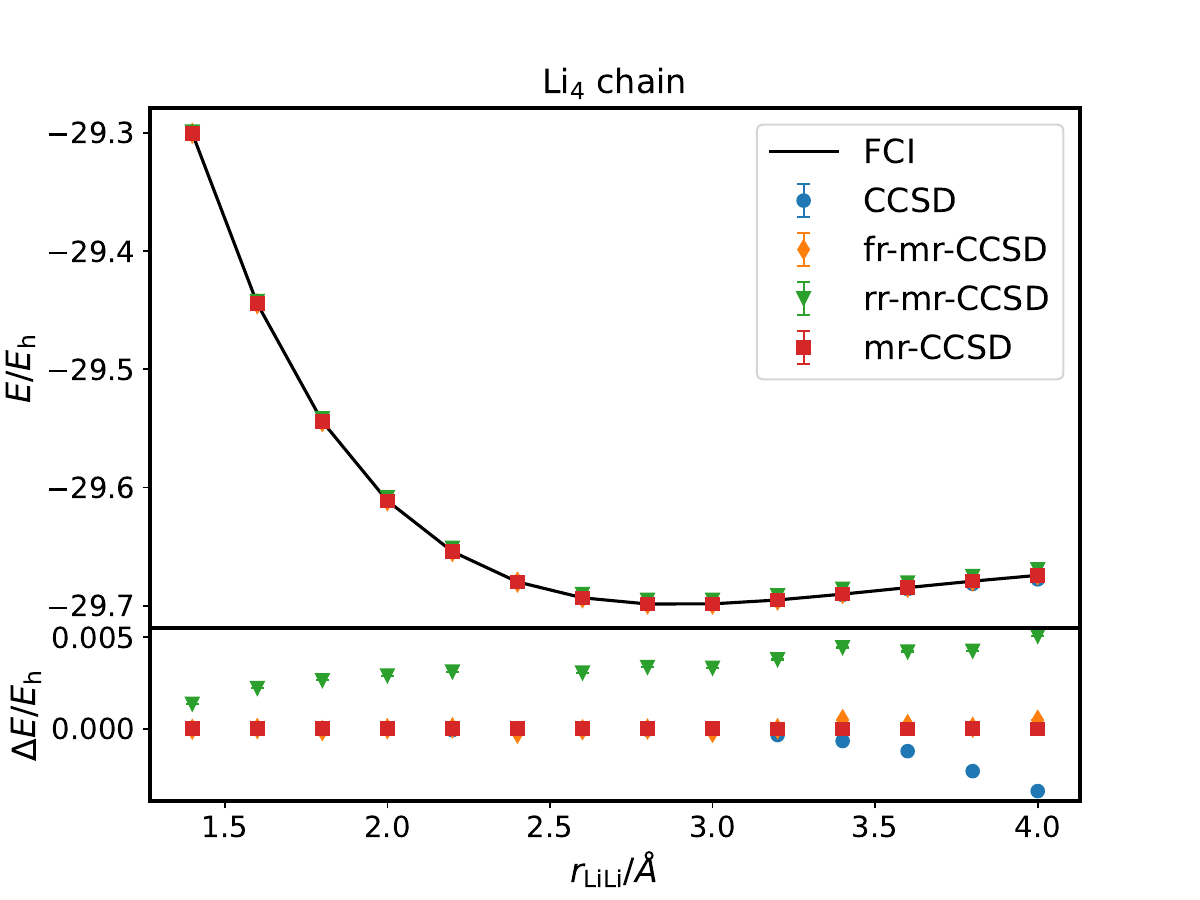}
\includegraphics[width =0.45\textwidth]{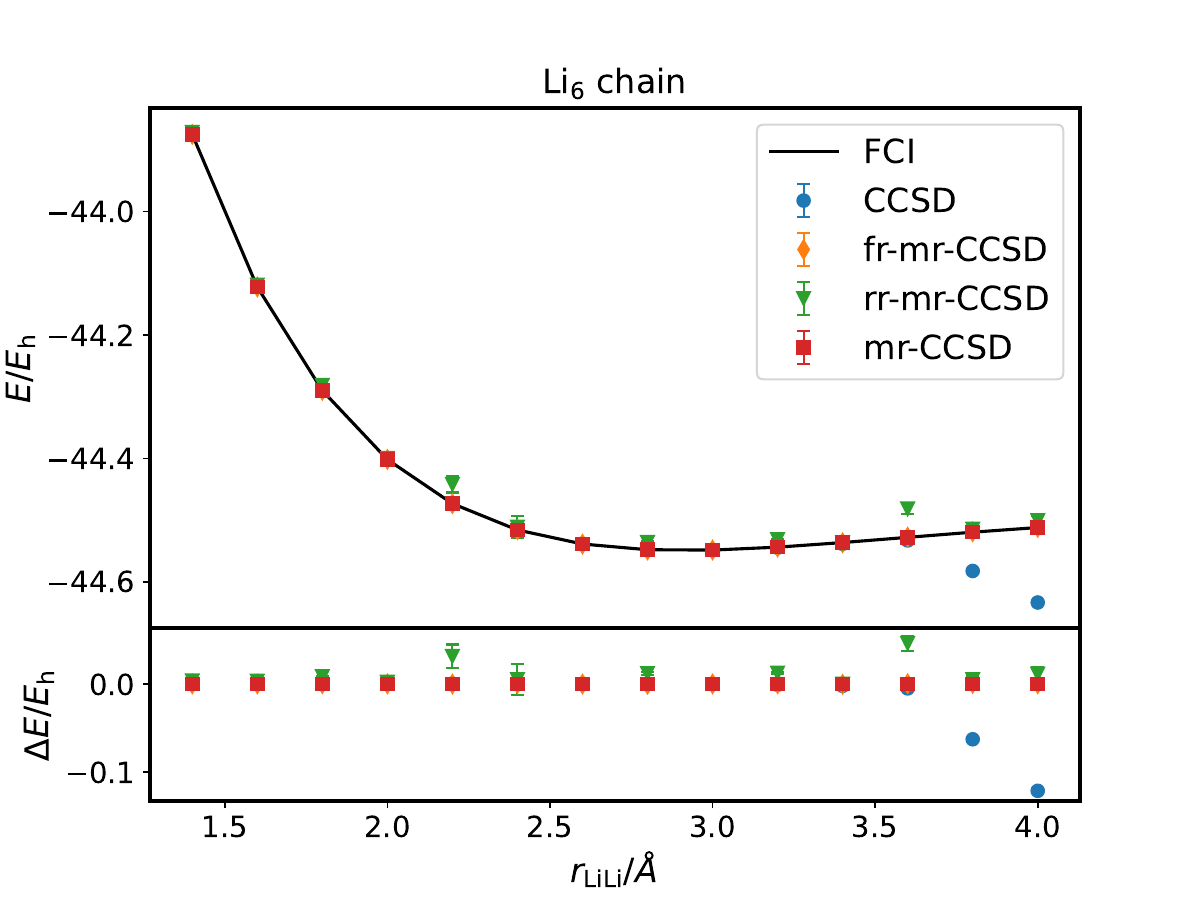}
\\
\includegraphics[width =0.45\textwidth]{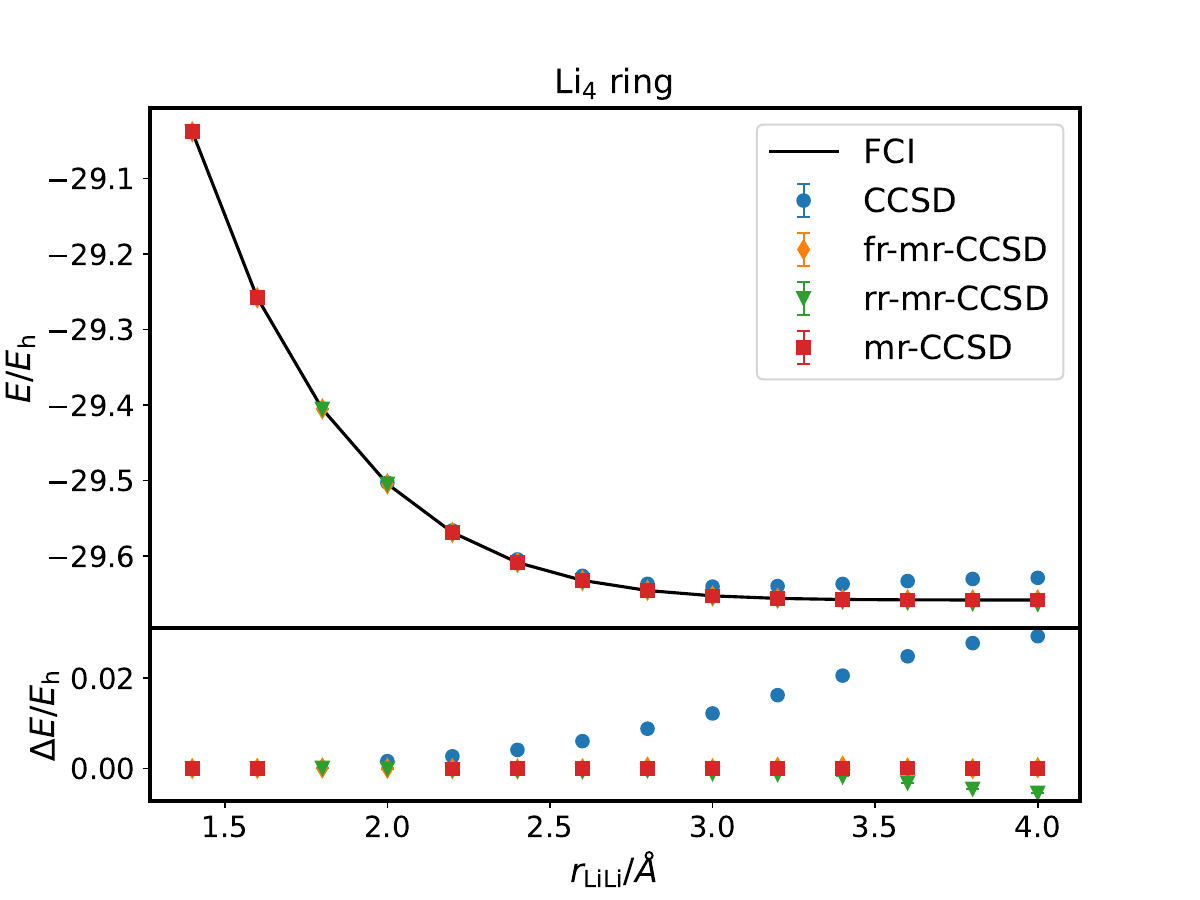}
\includegraphics[width =0.45\textwidth]{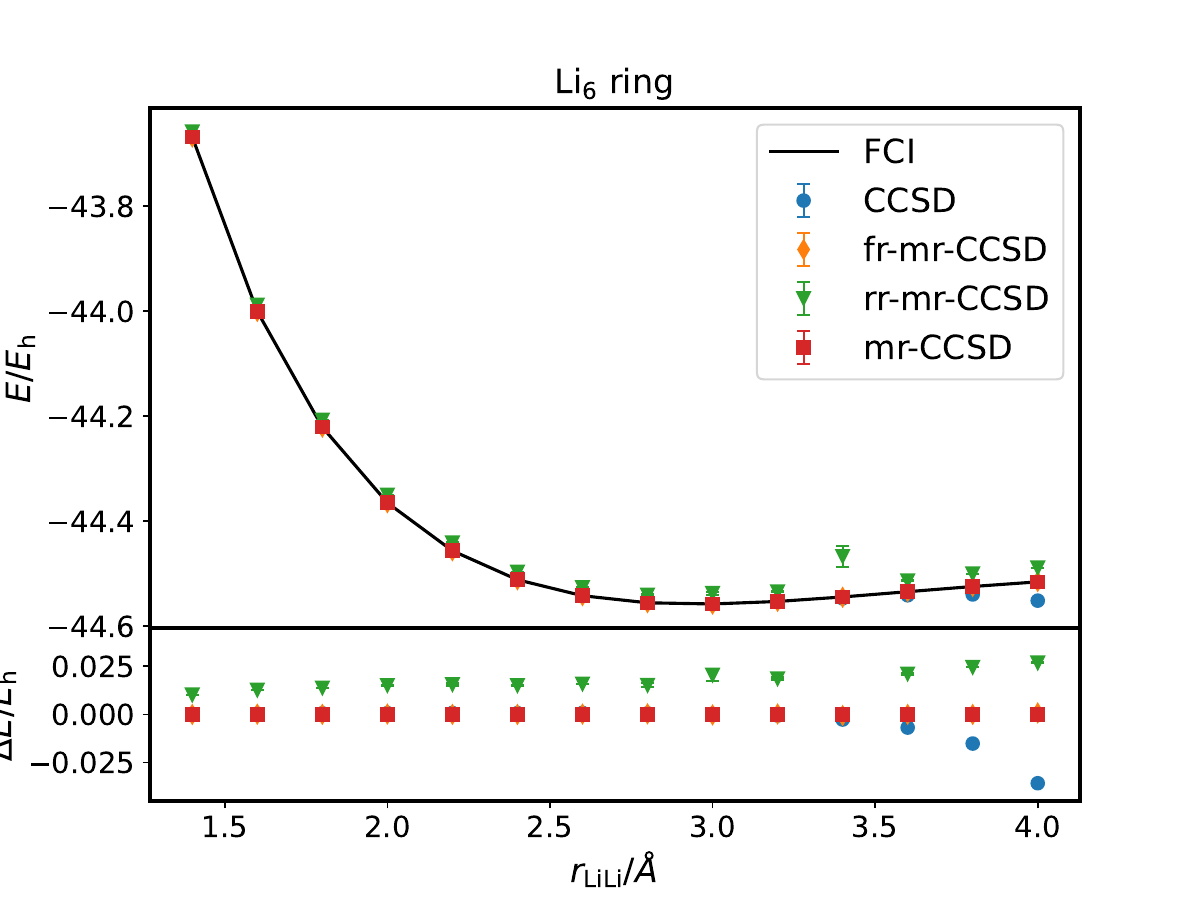}
\caption {\protect \footnotesize \raggedright Binding curves and energy errors computed with mr-CCMC, 
	fr-mr-CCMC and rr-mr-CCMC for the Li$_4$ chain (top left), Li$_4$ ring (bottom left), Li$_6$ chain (top right) and Li$_6$ ring (bottom right) with two $s$ orbitals per atom. 
	fr-mr-CCMCSD agrees with the exact mr-CCMCSD calculation in all cases, but the rr-mr-CCMCSD approach displays a systematic error in the Li$_4$ chain and Li$_6$ ring, as well as generating 
	a outliers in the energy curves of other systems.\\}
\label{fig:Li-struct} 
\end{figure*}

\begin{figure*}
\includegraphics[width =0.45\textwidth]{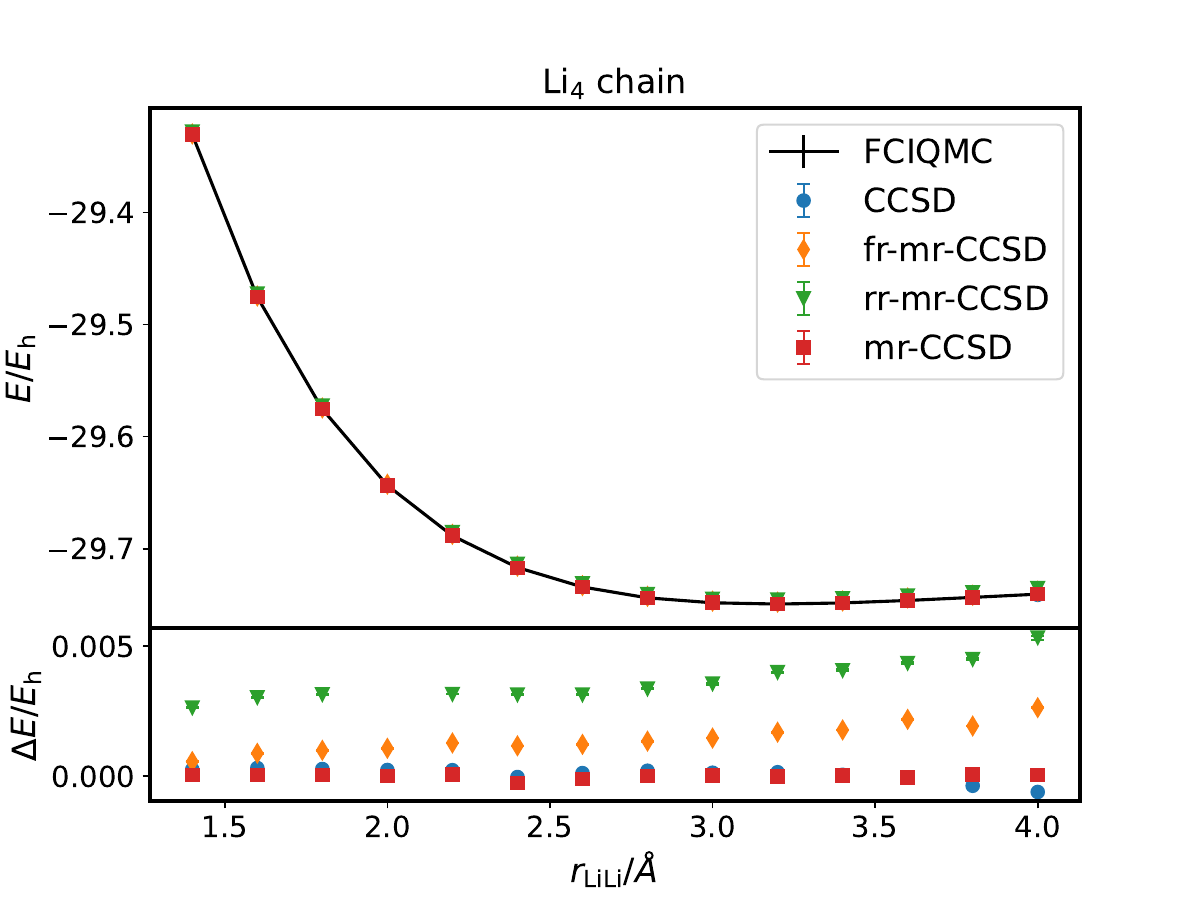}
\includegraphics[width =0.45\textwidth]{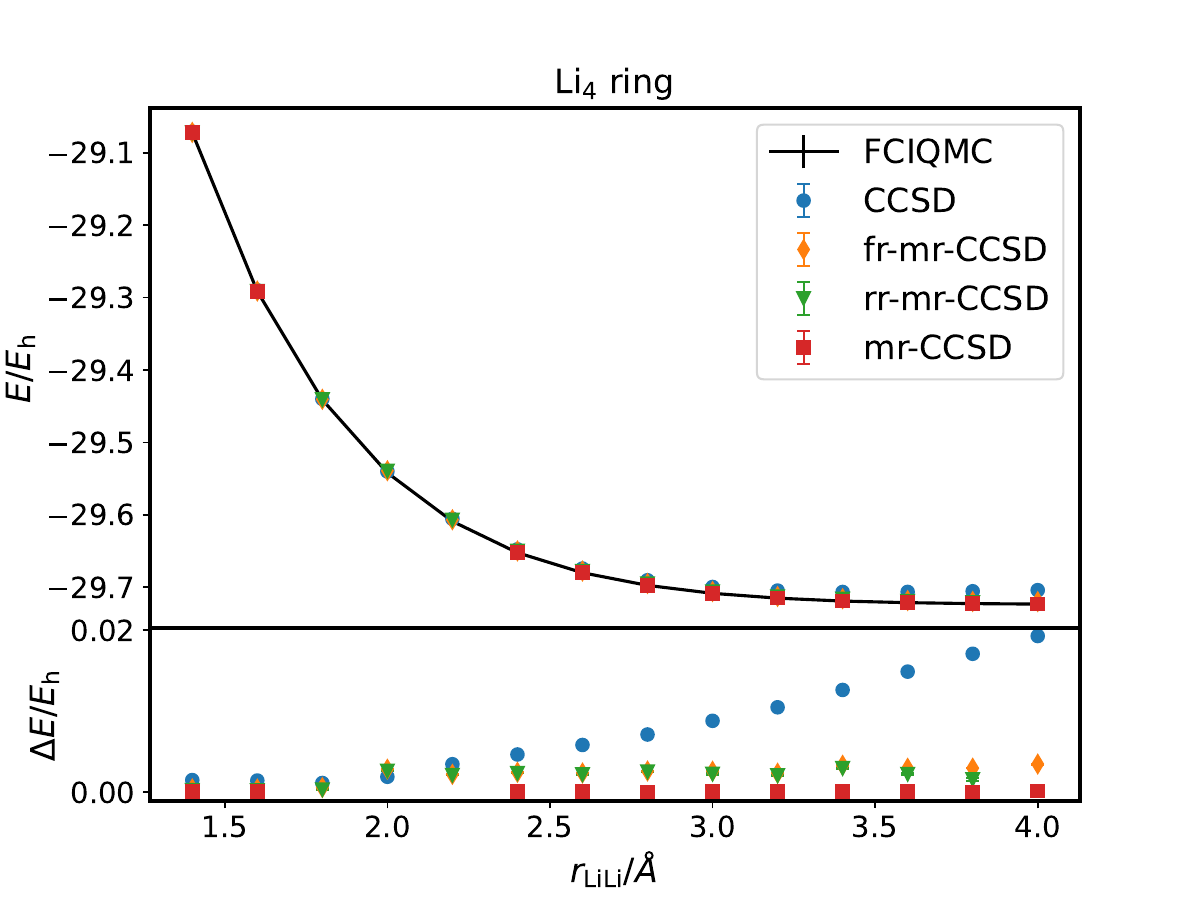}
	\caption[Binding curves and energy errors for Li$_4$ with three $s$ orbitals/atom.] {\protect \footnotesize \raggedright Binding curves and energy errors computed with mr-CCMC, fr-mr-CCMC and rr-mr-CCMC 
	for the Li$_4$ chain (left), Li$_4$ ring (right) with 3 $s$ orbitals per atom. FCI benchmark is obtained from FCIQMC. 
	For the ring structure, fr- and rr-mr-CCMC significantly outperform single-reference CCSD, but no longer
	agree perfectly with mr-CCMCSD. For the chain, CCMCSD itself is almost exact in the range of bondlengths studies, but the fr- and rr-mr-CCMC both now exhibit noticeable errors.}
	\label{fig:Li-struct2} 
\end{figure*}

Finally, we consider a set of small Li clusters, using reduced basis sets in which each Li atom has only two or three $s$ orbitals. 
We investigate 4- and 6-membered Li rings and chains over a wide range of Li-Li bond lengths. Near equilibrium, Li chains are well-behaved
single-reference systems, as is the 6-membered ring. The 4-membered Li ring has significant multireference character at all geometries due
to the presence of a partially occupied degenerate pair of spatial orbitals. Binding curves computed with FCI(QMC), 
mr-CCMCSD and partially relaxed approximations thereof are given in \cref{fig:Li-struct,fig:Li-struct2}. For the smaller of the two
basis sets, CCMCSD results start diverging from the FCI energy at large Li-Li separations. In comparison, mr-CCMCSD and
its approximations generally maintain excellent agreement with the exact results, with the notable exception of rr-mr-CCSD in the 
4-membered Li ring, which has noticeable errors (up to 5$\mathrm{m}E_\mathrm{h}$) across the entire binding curve. rr-mr-CCSD also
shows a number of outliers across the other binding curves, converging to the wrong energy values with significantly larger error
bars than neighbouring values. For the Li$_4$ ring, the mr-CCSD method is difficult to converge around $r_\mathrm{LiLi} = 1.9$\AA,
but both the frozen- and relaxed-reference approaches converge without issue over the entire binding curve. 

For the larger basis set, both CCSD and mr-CCSD are almost exact for the \ce{Li4} chain in the range of basis states considered. However, both the frozen- and relaxed-reference
approximations now have noticeable systematic errors. In contrast, for the \ce{Li4} ring, where CCMCSD diverges from FCI as $r_\mathrm{LiLi}$ increases, approximate mr-CCMCSD 
methods perform significantly better than the single-reference approach and converge over the whole range of bond-lengths, unlike the unapproximated mr-CCMCSD method.

\section{Conclusions}
Using the reference CASCIQMC wavefunction as a starting point, we develop two approximations to the mr-CCMC method. Frozen-reference mr-CCMC keeps the reference wavefunction fixed to its initial 
CASCIQMC value, while relaxed-reference mr-CCMC only allows spawning onto reference determinants from the external space. These two methods are shown to have different propagation requirements 
from the original algorithm. 
 In fr-mr-CCMC, maintaining a steady-state population can no longer be used  as a condition to find the shift as a wide range of values of $S$ give rise to stable 
populations. In order to get physically meaningful results, the shift is constrained to track the instantaneous projected energy, leading to the loss of one of the independent energy estimators 
from a conventional QMC calculation.

In rr-mr-CCMC, na\"ively spawning onto determinants in the reference space leads to a propagator that fails to converge 
to the ground state of the system even when the method is formally exact and is found to overestimate the correlation energy. 
This can be mitigated by constraining spawns to the space orthogonal to the CASCIQMC ground state. This approach is applied 
successfully to the Hubbard model, where the additional relaxation leads to better energy estimates than the frozen-reference approximation.
However, the fully stochastic implementation displays systematic errors when applied to a Li$_4$ chain and shows a propensity for 
spuriously converging to non-physical solutions in other Li structures as well. As noted in the deterministic examples, propagation
that is not fully orthogonal to the ground CASCI wavefunction leads to unphysical rr-mrCC energies. rr-mr-CCMC calculations are based on
a stochastic snapshot of the CASCI wavefunction which, particularly for relatively small walker numbers, is unlikely to be perfectly aligned with
the true CASCI ground state, so the random states generated during the propagation will not be exactly orthogonal to the ground state either, leading to
errors in the rr-mr-CCMC propagation.

Generally, these approximations are found to converge faster than the exact mr-CCMC method, as expected since they are initialised in a
multireference state, rather than with the HF determinant. While all 
examples in this paper use a CAS reference, this is not a requirement 
for any of the methods presented. Stochastic noise in the 
calculations is also reduced relative to the exact
approach, which allows the approximations to converge in regimes 
where mr-CCMC is ill-behaved. For the systems we have studied, 
allowing partial relaxation of the reference wavefunction does not 
normally lead to substantial accuracy gains and in some cases the 
method is more error-prone than the simpler frozen-reference a
pproximation. In cases where there is a significant difference 
between single- and multireference results, this  approximation 
generally significantly outperforms the single-reference approach, 
presenting a viable alternative when exact mr-CCMC calculations
would be unstable or prohibitively expensive.

{\section*{Supplementary Material}
See supplementary material for the basis set used for H$_4$ .}

\begin{acknowledgements} 
M-A.F. is grateful to the Cambridge Trust and Corpus
Christi College for a studentship and A.J.W.T. to the Royal Society for a
University Research Fellowship under Grant No. UF160398. 
\end{acknowledgements}
\section*{Data Availability Statement}
The data that support the findings of
this study are openly available in
the Apollo University of Cambridge Repository  at
http://doi.org/10.17863/CAM.93604, reference number
 6321755.

\section*{Author Declarations}
The authors have no conflicts to disclose.

\bibliography{./embedded}

\end{document}